\documentclass[12pt]{article}
\usepackage{amsfonts}
\usepackage{amssymb}
\usepackage{latexsym}
\usepackage{color}
\usepackage[active]{srcltx}
\newtheorem{theorem}{Theorem}

\newtheorem{corollary}{Corollary}
\newtheorem{conjecture}{Conjecture}

\newcommand{\pa}{\partial}

\newcommand{\om}{\omega}
\newcommand{\Om}{\Omega}

\newcommand{\De}{\Delta}

\newcommand{\rar}{\rightarrow}

\newcommand{\non}{\nonumber}

\newcommand{\be}{\begin{equation}}
\newcommand{\ee}{\end{equation}}
\newcommand{\ba}{\begin{array}}
\newcommand{\ea}{\end{array}}
\newcommand{\bea}{\begin{eqnarray}}
\newcommand{\eea}{\end{eqnarray}}
\newcommand{\bi}{\begin{itemize}}
\newcommand{\ei}{\end{itemize}}

\begin{document}

\title{The quantum n-body problem in dimension $d\ge n-1$: ground state }

\author{
     Willard Miller, Jr.\\[8pt]
School of Mathematics, University of Minnesota, \\
Minneapolis, Minnesota, MN 55455, U.S.A.\\[8pt]
miller@ima.umn.edu\\
[10pt]
Alexander V Turbiner\\[8pt]
Instituto de Ciencias Nucleares, UNAM, M\'exico DF 04510, Mexico\\[8pt]
turbiner@nucleares.unam.mx\\[8pt]
and \\[10pt]
M.A.~Escobar-Ruiz,\\[8pt]
Centre de Recherches Math\'ematiques, Universit\'e de Montreal, \\
C.P. 6128, succ. Centre-Ville, Montr\'eal, QC H3C 3J7, Canada\\[8pt]
escobarr@crm.umontreal.ca}

\date{\today}
\maketitle

\begin{abstract}
We employ generalized Euler coordinates for the $n$ body system in $d \geq n-1$ dimensional space, which consists of the centre-of-mass vector, relative (mutual), mass-independent distances $r_{ij}$ and angles as remaining coordinates. We prove that the kinetic energy of the quantum $n$-body problem for $d \geq n-1$ can be written as the sum of three terms: 
(i) kinetic energy of centre-of-mass, (ii) the second order differential operator $\De_{rad}$ which depends on relative distances alone and (iii) the differential operator $\Om$ which annihilates any angle-independent function. The operator $\De_{rad}$ has a large reflection symmetry group $Z_2^{\oplus \frac{n(n-1)}{2}}$ and in $\rho_{ij}=r_{ij}^2$ variables is an algebraic operator, which can be written in terms of generators of their {\it hidden} algebra $sl(\frac{n(n-1)}{2}+1, R)$. Thus, $\De_{rad}$ makes sense of the Hamiltonian of a quantum Euler-Arnold $sl(\frac{n(n-1)}{2}+1, R)$
top in a constant magnetic field. It is conjectured that for any $n$, the  similarity-transformed $\De_{rad}$ is the Laplace-Beltrami operator plus (effective) potential; thus, it describes a $\frac{n(n-1)}{2}$-dimensional
quantum particle in curved space. This  was verified for $n=2,3,4$.
After de-quantization the similarity-transformed $\De_{rad}$ becomes the Hamiltonian of the
classical top with variable tensor of inertia in an external potential.

This approach allows a reduction of the $dn$-dimensional spectral problem to a $\frac{n(n-1)}{2}$ -dimensional spectral problem if the eigenfunctions  depend only on relative distances. We prove  that the ground state function of the $n$ body problem depends on relative distances alone.

\end{abstract}


\vskip 2cm

\section{Introduction}
As a continuation of \cite{paper1},\cite{paper2}, we study the Hamiltonian for an $n$-body quantum system of $d$-dimensional massive  particles
(of $d$ degrees of freedom), with translation-invariant potential that depends on relative distances
between particles only. Thus the Hamiltonian is defined on an $(n\,d)$-dimensional configuration space.
The number of relative distances between particles is  $\frac{n(n-1)}{2}$. If  $d\ge n-1$,
the relative distances are edges of an  $n$-vertex polytope   and they are functionally-independent. Thus, they can be used as new independent variables.
Separation of the center of mass accounts for $d$ variables while the remaining $\frac{(n-1)(2d-n)}{2}$ variables are angular. The number of angular variables is equal to the dimension of the homogeneous space ${SO(d)}/{SO(d-n+1)}$. It corresponds to rotation of the $n$-vertex regular polytope of interaction in
$\frac{(n-1)(2d-n)}{2}$ dimensional space.

If the number of particles is $n > d+1$\,, the dimension of the space of relative distances is  $\frac{d(2n-d-1)}{2}$. This implies the existence of constraints between relative distances.
The polytope of interaction becomes degenerate: a number of relative distances are diagonals,
rather than edges. In this case $\frac{(n-d)(n-d-1)}{2}$ diagonals are functionally-dependent.
The number of angular variables is equal to  $\dim SO(d)= \frac{d(d-1)}{2}$.
In particular, for one-dimensional particles when $d=1$, corresponding to $n$ interacting particles
on the line, only $(n-1)$ relative distances are needed  to parameterize the relative motion; angular variables are absent. For two-dimensional particles when $d=2$, corresponding to $n$ interacting
particles on the plane, $(n - 3)$, relative distances (or, differently speaking, {\it radial} variables)
are needed while there is a single angular variable; the system is $SO(2)$-rotationally-invariant.
For the  physically important case of three-dimensional particles, $d=3$, $(3n - 6 + \delta_{2,n})$
radial variables are needed; for $n>2$ there are three angular variables and the system is $SO(3)$-rotationally-invariant.

Here we focus on the case $d \ge n-1$, so that there are no constraints between relative distances.
Then in appropriate new coordinates the Hamiltonian restricted  to the $(n-1)\times d$ dimensional space
of relative motion is the flat-space Laplace operator plus potential.  We prove explicitly that
the Laplace operator splits into a sum of two second-order differential operators where
the first operator depends on relative distances {\it only} and the second operator depends on
angular derivatives in such a way that it annihilates any angle-independent function.
For $S$-states, those for which  the total angular momentum of the system is zero and
eigenfunctions of the Hamiltonian have no angular dependence, only the first radial operator
is relevant. We compute this operator explicitly and show that because of the existence of a hidden algebra the Hamiltonian is exactly-solvable. The choice of angular variables is immaterial to the final result.

\section{$n$-body problem in dimension $d>n-2$, S-states}

The  Hamiltonian for the $n$-body quantum system of $d$-dimensional particles with translation-invariant potential, depending on relative distances between particles only, is of the form,
\begin{equation}
\label{Hgen}
   {\cal H}\ =\ -\sum_{i=1}^n \frac{1}{2m_i}\De_i^{(d)}\ +\  V(r_{jk})\ ,\
\end{equation}
with coordinate vector of $i$th particle ${\bf r}_i \equiv {\bf r}^{(d)}_i=(x_{i,1}\,,\cdots \,,x_{i,d})$\,, where
\begin{equation}
\label{rel-coord}
r_{jk}=|{\bf r}_j - {\bf r}_k|\ ,\
\end{equation}
is the (relative) distance between particles $j$ and $k$ and $m_i$ is the mass of particle $i$.
The number of relative distances is equal to the number of edges of the polytope
formed by taking the body positions as vertices. We call this  the {\it polytope of interaction}. Here, $\De_i^{(d)}$ is the $d$-dimensional Laplacian,
\[
     \De_i^{(d)}\ =\ \frac{\pa^2}{\pa{{\bf r}_i} \pa{{\bf r}_i}}\ ,
\]
associated with the $i$th body. The configuration space for ${\cal H}$ is ${\bf R}^{n \times d}$.
The center-of-mass motion described by vectorial coordinate
\[
    {\bf R}_{_0} \ =\ \frac{1}{\sqrt{M_n}}\,\sum_{k=1}^{n} m_k {\bf r}_{_k}\ ,\ M_n={\sum_{j=1}^nm_j}\ ,
\]
can be separated out; this motion is described by a $d$-dimensional plane wave, here $M_n$ is total mass of the system.

The spectral problem is formulated in the space of relative motion
${\bf R}_r \equiv {\bf R}^{d(n-1) }$; it is of the form,
\begin{equation}
\label{Hrel}
   {\cal H}_r\,\Psi(x)\ \equiv \ \bigg(- \De_r^{(d(n-1))} + V(r_{jk})\bigg)\, \Psi(x)\ =\ E \Psi(x)\ ,\
   \Psi \in L_2 ({\bf R}_r)\ ,
\end{equation}
where $\De_r^{(d(n-1))}$ is the flat-space Laplacian in the space of relative motion.. Let $M_j=\sum_{k=1}^jm_k$, $j=1,\cdots,n-1$. If the space of relative motion ${\bf R}_r$ is parameterized by $(n-1)$, $d$-dimensional vectorial Jacobi coordinates
\be\label{Jacobi}
     {\bf r}^{(F)}_{j} \ = \ \sqrt{\frac{m_{j+1}M_j}{M_{j+1}}}\left({\bf r}_{j+1}-\sum_{k=1}^j\frac{m_k{\bf r}_k}{M_j}\right) ,
        \qquad\qquad j=1,\cdots,n-1 ,
\ee
the flat-space, $(dn)$-dimensional Laplacian in the space of relative motion becomes diagonal, e.g. \cite{Gu}, and the original Hamiltonian we started with takes the form
\begin{equation}
\label{Dflat}
   {\cal H}_0=    \sum_{i=1}^n\frac{1}{m_i}\De_i^{(d)} \ = \ \De_{{\bf R}_0}\  +  \
       \sum_{i=1}^{n-1}\frac{\pa^2}{\pa{{\bf r}_i^{(F)}} \pa{{\bf r}_i^{(F)}}}\ .
\end{equation}

Again we  choose new coordinates for the problem, divided into ``radial coordinates" (dimensionful) and ``angular coordinates" (dimensionless).
Considering  ${\bf r}^{(F)}_{j}$ as a column vector, we write
\be
\label{Coords}
{\bf r}^{(F)}_{1}=O\left(\ba{c} 0\\ \vdots\\ \vdots\\ \vdots\\ 0\\a_1\ea\right),\
{\bf r}^{(F)}_{2}=O\left(\ba{c} 0\\ \vdots\\ \vdots\\  0\\a_3\\a_2\ea\right),\ \cdots,\
{\bf r}^{(F)}_{n-1}=O\left(\ba{c} 0\\ \vdots\\ 0\\  a_N\\ a_{N-1}\\ \vdots \\ a_{S}\ea\right)
\ee
where $N=\frac{n(n-1)}{2}$, $S=\frac{n^2-3n+4}{2}$, and we can assume that $a_N\ge 0$. Here, $O\in SO(d)/G$ where
\be\label{Gdefa1} G=\left\{{\tilde O}\in SO(d): {\tilde O}\left(\ba{c} 0\\ \vdots\\ \vdots\\ \vdots\\
0\\a_1\ea\right)=\left(\ba{c} 0\\ \vdots\\ \vdots\\ \vdots\\ 0\\a_1\ea\right),  \cdots,
{\tilde O}\left(\ba{c} 0\\ \vdots\\ 0\\  a_N\\ a_{N-1}\\ \vdots \\ a_{n-1}\ea\right)=\left(\ba{c} 0\\ \vdots\\ 0\\  a_N\\ a_{N-1}\\ \vdots \\ a_{S}\ea\right)
\right\},\ee
and we restrict the $\{ {\bf r}^{(F)}_{k}\}$ to the domain where they form a linearly independent set.
The number of angular coordinates $\phi_k$ where $O=O(\phi_k)$ is
\[ \dim SO(d)/SO(d-n+1)=\frac{d(d-1)}{2}-\frac{(d-n+1)(d-n)}{2}=\frac{(n-1)(2d-n)}{2}.\]
and the number of radial variables $a_\ell$ is $\frac{n(n-1)}{2}$.

There are $\frac{n(n-1)}{2}$ inner products  $\{{\bf r}^{(F)}_{k}\cdot  {\bf r}^{(F)}_{\ell}\}$, independent of the angular variables and expressible
as quadratic functions of the radial variables.
The vector ${\bf r}^{(F)}_{k}$ has coefficients
\be\label{rdef} {\bf r}^{(F)}_{k}|_j\equiv {y}_j^k=\sum_{\ell=\frac{k(k-1)}{2}+1}^{\frac{k(k+1)}{2}}a_\ell\,
O_{j,d-\ell+\frac{k(k-1)}{2}+1},\quad j=1,\cdots,d,\ee
and the quadratic relations are, for $k\le\ell$,
\be\label{quadrelations}\sum_{j=1}^d y_j^ky_j^\ell=\sum_{h=0}^{k-1} a_{\frac{k^2-k+2}{2}+h}a_{\frac{\ell^2-\ell+2}{2}+h}.\ee

\begin{theorem} The Laplacian $\sum_{i=1}^n\frac{1}{2m_i} \De_i^{(d)}$ splits as
\[
 \sum_{i=1}^n\frac{1}{2m_i} \De_i^{(d)}\ =\  \De_{{\bf R}_0}
 \ +\ \De_{\rm rad}\ +\ \Om \ ,
\]
where $\De_{{\bf R}_0}$ is the center of mass Laplacian, $\De_{\rm rad}$ depends on the radial variables ${\bf a}$ alone, and $\Om$ annihilates any function of the radial variables alone. The first two operators are stable, they do not depend on the choice of angular variables
 \end{theorem}

 \medskip\noindent {\bf Proof}:
Differentiating (\ref{quadrelations}) we obtain
\be
\label{quadrelations2}
 y_j^\ell\ =\ \sum_{h=0}^{k-1}\left( \frac{\pa a_{\frac{k^2-k+2}{2}+h}}{\pa y_j^k}a_{\frac{\ell^2-\ell+2}{2}+h}
+a_{\frac{k^2-k+2}{2}+h}\frac{\pa a_{\frac{\ell^2-\ell+2}{2}+h}}{\pa y_j^k}\right)\ .
\ee

In principle we  can solve (\ref{quadrelations2}) for the $\frac{\partial a_t}{\partial y_j^p}$, recursively, first for $\ell=k, k=1,2,\cdots$
and then for $\ell=k+1,k+2,\cdots$. An important thing to notice is that at each step in the recursion each nonzero
$\frac{\partial a_t}{\partial y_j^p}$, is a  sum of terms, each term of which is a rational function of the radial variables alone
times $y_j^h$ for some $h$ and this last factor contains the dependence on the angular variables.
 Then
from the chain rule and (\ref{rdef}), (\ref{quadrelations}),   we can verify that the partial derivatives $\partial_{{ y}_j^k}$ take the form
\be\label{xjderb} \partial_{{ y}_j^k}=\sum_{\ell=1}^NS^k_j({\bf a},O,\ell)\partial_{a_\ell}+\Omega_j^{(k)},\ee
where \be\label{Tident} S_j^k({\bf a},O,\ell)=\sum_{s=1}^NT({\bf a},\ell,k)_s\, O_{j,d-s+1},\ee
 the $T({\bf a},\ell,k)_s$ are simple rational functions of the radial variables alone,
 \be\label{Omegaopb} \Omega^{(k)}_j=\sum_{\ell=1}^{\frac{(n-1)(2d-n)}{2}}\frac{\partial \phi_\ell}{\partial { y}_j^k}\partial_{\phi_\ell},\ee
 and the $\phi_\ell$ are the angular variables.
To verify the theorem we need first to show that
\[ \sum_{j=1}^d S_j^k({\bf a},O,\ell)^2\]
and
\[ \sum_{j=1}^d S_j^k({\bf a},O,\ell)\frac{\partial S_j^k({\bf a},O,\ell)}{\partial a_\ell}\]
are both functions of the radial parameters $\bf a$ alone, for each $k,\ell,O$. But this follows immediately from (\ref{Tident}) and the orthogonality of matrix $O$.

Finally, we need to show that $\sum_{j=1}^d \Om^{(k)}_jO_{j,d-\ell}$ is a function of the radial variables alone for each $k,\ell,O$.
 Although we don't have  explicit expressions for the operators (\ref{Omegaopb}) we can use expressions (\ref{xjderb}) to define
their actions. Indeed, applying both sides of (\ref{xjderb}) to ${y}_j^h$ we find
\[ \delta_{kh}= \sum_{\ell=1}^N S_j^k({\bf a},O,\ell)\,O_{j,d-\ell+\frac{k(k-1)}{2}+1}+\sum_{s=\frac{h(h-1)}{2}+1}^{\frac{h(h+1)}{2}}a_s\Omega_j^{(k)}O_{j,d-s+\frac{h(h-1)}{2}+1}
\]
Summing on $j$ and using the
fact that $O$ is an orthogonal matrix, we
obtain
\[ d\,\delta_{kh}= \sum_{\ell=\frac{k(k-1)}{2}+1}^N T({\bf a},\ell,k)_{\ell-\frac{k(k-1)}{2}}  +
\sum_{s=\frac{h(h-1)}{2}+1}^{\frac{h(h+1)}{2}}a_s\,\left(\sum_{j=1}^d\Omega_j^{(k)} O_{j,d-s+\frac{h(h-1)}{2}+1}\right). \]
Fixing $k$ and choosing $h=1,2,\cdots,n-1$ in that order, we can solve this equation recursively for each of
\[ H^k_\ell({\bf a})=\left(\sum_{j=1}^d\Omega_j^{(k)} O_{j,d-\ell}\right),\quad \ell=1,2,\cdots,N\]
as functions of the radial variables alone.

From these results it is easy to show that  the Laplacian splits in the form
\begin{equation}\label{splithamiltonian}
  \sum_{i=1}^{n-1} \frac{\pa^2}{\pa{{\bf r}_i^{(F)}} \pa{{\bf r}_i^{(F)}}}\ =\
  \Delta_{\rm rad}+\Omega \ ,
\end{equation}
where the operator $\Delta_{\rm rad}$ depends on the radial variables only and the operator $\Omega$
annihilates any function of the radial variables alone. \quad
$\Box$

\begin{corollary}
The reduced Hamiltonian $ { H}_{\rm rad} \ =\ -\De_{\rm rad}\ +\ V$ admits the symmetry algebra $so(n-1)$ of 1st order symmetry (differential) operators.
\end{corollary}

\medskip\noindent {\bf Proof}: The Hamiltonian
$-\frac{\pa^2}{\pa{{\bf r}_i^{(F)}} \pa{{\bf r}_i^{(F)}}}\ +V(r_{jk})$
admits the symmetry algebra

\noindent
$so(d(n-1))$ with generators $y_{j_1}^{k_1}\partial_{y_{j_2}^{k_2}}-y_{j_2}^{k_2}\partial_{y_{j_1}^{k_1}}$. These are not symmetries of the reduced Hamiltonian, however. Now consider the subset of symmetries
\[{\tilde K}_{jk}=\sum_{\ell=1}^d \left(y_{j}^{\ell}\partial_{y_{k}^{\ell}}-y_{k}^{\ell}\partial_{y_{j}^{\ell}}\right), \quad 1\le j<k\le n-1.\] Clearly, the $\{ {\tilde K}_{jk}\}$ generate the algebra $so(n-1)$.
Using the relations (\ref{rdef}), (\ref{xjderb}), (\ref{Tident}) and the orthogonality of the matrix $O$, it follows easily that the symmetries decompose as
\[  {\tilde K}_{jk}=K_{jk} +\Omega_{jk},\]
where $K_{jk}$ depends on the radial variables $\bf a$ alone and the operators $\Om_{jk}$ annihilate any function of the radial variables alone. Thus, the $\frac{(n-1)(n-2)}{2}$ operators  $\{ K_{jk}\}$ generate the algebra $so(n-1)$ and satisfy $[ H_{\rm rad},K_{jk}]=0$. \quad $\Box$

The existence of this symmetry algebra permits us to separate some variables from the $S$-state,
reduced eigenvalue problem, see below (\ref{redrad}). Indeed, the  basis of symmetry operators
for a maximal Abelian subalgebra of $so(n-1)$ is mutually commutative, so can be simultaneously diagonalized.
The number of separable variables so obtained is equal to the dimension of the  subalgebra,
i.e. to the rank of $so(n-1)$, \cite{Freudenthal} This rank equals $\frac{n-2}{2}$ for $n$ even, and $\frac{n-1}{2}$
for $n$ odd. Thus, we can separate one variable for n=3,4, two variables for n=5,6, etc.
We {\it conjecture} that this is the maximum number of variables that can be separated.

By construction, the $n$-body Hamiltonian (\ref{Hgen}) is essentially self-adjoint with respect to the volume
measure $dv=\prod_{1\le j\le n\, 1\le k\le d}dy_j^k$.
Moreover,
by expanding  the Jacobian determinant by minors, repeatedly, for the change of variables (\ref{xjdera}),(\ref{yjdera}),
one can verify that the transformed volume measure takes the form
\[
dv\ =\ \prod_{j=2}^n a_{j(j-1)/2}^{d-j+1}\, da_1\ da_2\cdots da_{n(n-1)/2}\ d\Om
\]
where $d\Om$ is the angular part of the measure. Thus, the volume measure factors
 to the product of radial and angular measures.
It follows that the radial reduced Hamiltonian $(-\De_{\rm rad}+V)$ is essentially
self-adjoint with respect to the radial measure
\be
\label{vrad-A}
      dv_{\rm rad}\ =\ \prod_{j=2}^n a_{j(j-1)/2}^{d-j+1}\,da_1\ da_2\cdots da_{n(n-1)/2}\ .
\ee

Now we turn our attention to the explicit computation of $\De_{\rm rad}$. First we introduce new coordinates.
Note that
\be
\label{rFr}
   {\bf r}_j-{\bf r}_\ell=\sqrt{\frac{M_j}{m_jM_{j-1}}}\,{\bf r}_{j-1}^{(F)}\ +\  \sum_{s=\ell}^{j-2}\sqrt{\frac{m_{i+1}}{M_iM_{i+1}}}{\bf r}_{s}^{(F)}
   -\sqrt{\frac{M_{\ell-1}}{m_\ell M_\ell}}\,{\bf r}_{\ell-1}^{(F)}
\ee
for $j>\ell\ge 1$, where we define ${\bf r}_0^{(F)}=0$, \cite{Gu}. Now introduce new radial coordinates
\be
\label{rhocoords}
 \rho_{j\ell}\ =\ \rho_{\ell j}\ =\ ||{\bf r}_j-{\bf r}_\ell||^2=({\bf r}_j-{\bf r}_\ell)\cdot ({\bf r}_j-{\bf r}_\ell)\ .
\ee
We see from (\ref{Coords}) and (\ref{rFr}) that  the $N$ distinct variables $\rho_{ij}$ are independent
of the angular variables and are quadratic functions of the $N$ radial variables $a_k$. Moreover,
it is easy to see geometrically that the $\rho$ variables are generically functionally independent.
Thus, they can be used as an alternate radial coordinate system.

\begin{theorem}
 For $S$-states of the $n-$body problem ($n \geq 2$) in arbitrary $(n \times d)$-space with $d \geq n-1$
 the differential part of the ``reduced radial Hamiltonian"
\be
\label{redrad}
  { H}_{\rm rad} \ =\ -\De_{\rm rad}\ +\ V
\ee
is algebraic in the $\rho_{ij}$ coordinates:
\begin{equation}
\label{eqn20}
\De_{\rm rad}(\rho)\ =\
2\,\sum_{i \neq j,i\neq k, j< k}^{n}\,\frac{1}{m_i}(\rho_{ij} + \rho_{ik} - \rho_{jk})\pa_{\rho_{ij}}\pa_{\rho_{ik}}\ +
\]
\[
2\,\sum_{i<j}^n \bigg(\frac{m_i+m_j}{m_i m_j} \bigg) \rho_{ij} \pa^2_{\rho_{ij}} +  d\, \sum_{i<j}^n\,\bigg(\frac{m_i+m_j}{m_i m_j}\bigg)\pa_{\rho_{ij}}
    \ .
\end{equation}

\end{theorem}

\medskip\noindent {\bf Proof}: We consider the Laplacian $\sum_{i=1}^n\frac{1}{2m_i} \De_i^{(d)}$, expressed in terms of the center of mass, radial and angular variables.
Here, we take the radial variables as
\[
     \rho_{\ell k}=\sum_{s=1}^d(x_{\ell,s}-x_{k,s})^2,\quad 1\le \ell<k\le n\ ,
\]
cf. (\ref{rhocoords}).
Thus,
\[
\pa_{x_{\ell,s}}=2\sum_{h\ne \ell,}(x_{\ell,s}-x_{h,s})\pa_{ \rho_{\ell h}} + \cdots \ ,
\]
where the non-explicit terms are  partial derivatives in the angular and center of mass variables. From this we can compute the coefficient of $\partial_{ \rho_{\ell k}}^2$
in $\sum_{i=1}^n\frac{1}{2m_i} \De_i^{(d)}$ for $\ell<k$. It is
\[
2 \bigg(\frac{1}{m_\ell}+\frac{1}{m_k}\bigg)\sum_{s=1}^d(x_{\ell,s}-x_{k,s})^2=
2\frac{m_\ell+m_k}{m_\ell m_k}\rho_{\ell k}\ .
\]
Similarly the coefficient of  $\pa_{ \rho_{\ell k}}\pa_{ \rho_{\ell {k'}}}$ for $k< k'$ is
\[
\frac{ 4}{m_\ell}\sum_{s=1}^d(x_{\ell,s}-x_{k,s})(x_{\ell,s}-x_{{k'},s})
=\frac{4 }{m_\ell}({\bf r}_\ell-{\bf r}_k)\cdot({\bf r}_\ell-{\bf r}_{k'})=\frac{2}{m_\ell}(\rho_{\ell {k}}+\rho_{\ell {k'}}-\rho_{k {k'}})\ ,
\]
where the last equality follows from the law of cosines. The coefficient of
$\pa_{ \rho_{\ell k}}\pa_{ \rho_{{\ell'} {k'}}}$ for $k,{k'},\ell,{\ell'}$ all
pairwise distinct is $0$. The coefficient of $\pa_{ \rho_{\ell k}}$ for $\ell<k$ is
\[
      d\, \bigg(\frac{1}{m_\ell}+\frac{1}{m_k}\bigg)\ .
\]
Thus, we have determined all terms in $\De_{\rm rad}$.\quad $\Box$

\begin{conjecture}
All symmetry operators $K_{ij}$ are algebraic operators in the $\rho$ variables with linear coefficients in $\rho$.
\end{conjecture}
It is certainly the case for $n=3$ and $n=4$, see below Examples 1-2.

Formula (\ref{eqn20}) is the main result of this paper. It has to be emphasized that it takes the amazingly simple form
of the operator (\ref{eqn20}).
In general, the operator $\De_{rad}$  can be written as
\begin{equation}
\label{gmunu}
     \De_{rad}\ =\ g^{\mu \nu} \pa_{\mu} \pa_{\nu}\ +\ b^{\mu} \pa_{\mu}\ ,
\end{equation}
where $g^{\mu \nu}$ is the matrix made out of coefficients in front of the second derivatives and $b^{\mu}$ is a
column vector. We make sense of $g^{\mu \nu}$ as the contravariant metric tensor, for a Riemannian space; in particular it is positive definite.
The easiest way to see this is to consider the kinetic energy for the flat space Hamiltonian (\ref{splithamiltonian}),
${\cal H}_0=-\sum_{i=1}^n\frac{1}{m_i}\De_i^{(d)}$.
In flat space coordinates it is clear that the kinetic energy is positive definite.
This means that for any choice of spatial coordinates $\{u_j\}$ the flat space contravariant tensor $G^{ij}({\bf u})$ has the property that the kinetic energy
$\sum_{ij}G^{ij}({\bf u})p_{u_i}p_{u_j}>0$ for any nonzero momentum vector $\{p_{u_j}\}$. Now let us choose (radial and
angular) spatial coordinates $\{\rho_\mu\equiv \rho_{ij},\phi_k\}$
as introduced above. In terms of these coordinates the kinetic energy satisfies the inequality
\begin{equation}
\label{inequality}
   \sum_{\mu,\nu}g^{\mu\nu}({ \rho})p_{\rho_\mu}p_{\rho_\nu}+2\sum_{\mu, k}g^{\mu k}
   ({\rho},{\phi}) p_{\rho_\mu}p_{\phi_k}
   + \sum_{j,k}g^{jk}({ \rho},{ \phi})p_{\phi_j}p_{\phi_k}>0 \ ,
\end{equation}
for {\it any} nonzero momentum vector $\{p_{\rho_\mu},p_{\phi_k}\}$. Now we restrict the momentum vector so that all angular terms
$p_{\phi_k}=0$. Then we have the inequality $ \sum_{\mu,\nu}g^{\mu\nu}({ \rho})p_{\rho_\mu}p_{\rho_\nu}>0$ for all nonzero vectors
$\{p_{\rho_\mu}\}$. This implies that $g^{\mu\nu}$ is a positive definite tensor determining a Riemannian manifold. The quantum argument is
similar. This means that the inner product
\[
  <\Psi,{\cal H}_0\Psi> =\int_{{\bf R}^{n\times d}} \sum_{ij} G^{ij}({\bf u})\,\pa_{u_i}\Psi\,
    \overline {\pa_{u_j}\Psi} \, dv>0 \ ,
\]
for all nonzero ${\cal C}^2 $ functions in the domain of ${\cal H}_0$, e.g. \cite{Hellwig}. Applying an argument analogous to (\ref{inequality})
we see that, upon restricting $\Psi$ to a nonzero function independent of the angular variables, we have
\[
  <\Psi,\De_{LB}(\rho)\Psi>=\int_{V_n>0} \sum_{\mu\nu}g^{\mu \nu}(\rho)\,\pa_{\rho_mu}\Psi\,
  \overline {\pa_{\rho_\nu}\Psi}\, d\om>0\ ,
\]
where $V_n$ is the volume of the $n$-polytope of interaction, $d\omega$ is the volume element for the Riemannian space and $\Delta_{LB}$ is the Laplace-Beltrami operator
on the Riemannian manifold.
(Recall that $dv=d\om({\rho})\, d\Omega(\phi)$ where the integral over the angular variables ${\phi }$ just yields a constant.)
Thus, the Laplace-Beltrami operator is positive definite.

\subsection{The representations of $sl(M+1,{\bf R})$}

The operator (\ref{eqn20}) is $sl(M+1,{\bf R})$-Lie algebraic - it can be rewritten in terms of the
generators of the maximal affine subalgebra $b_{M+1}$ of the algebra $sl(M+1,{\bf R})$, where $M=\frac{n(n-1)}{2}$, realized by the first order differential operators,
see \cite{Turbiner:1988,Turbiner:2016,RT:1995,Brink:1997}
\begin{eqnarray}
\label{slMR}
 {\cal J}_i^- &=& \frac{\pa}{\pa u_i}\ ,\qquad \quad i=1,2,3, \ldots M , \non  \\
 {{\cal J}_{ij}}^0 &=&
               u_i \frac{\pa}{\pa u_j}\ , \qquad i,j=1,2,3, \ldots M \ , \\
 {\cal J}^0(N) &=& \sum_{j=1}^{3} u_j \frac{\pa}{\pa u_j}-N\, , \non \\
\hskip -0.6cm {\cal J}_i^+(N) &=& u_i {\cal J}^0(N)\ =\
    u_i\, \left( \sum_{j=1}^{3} u_j\frac{\pa}{\pa u_j}-N \right)\, ,
       \, i=1,2,3, \ldots M\, ,
\end{eqnarray}
where $N$ is a parameter. Generators ${\cal J}_i^-, {{\cal J}_{ij}}^0$ span the algebra
$b_{M+1} \in sl(M+1,{\bf R})$. Generators ${\cal J}_i^-, {{\cal J}_{ij}}^0, {\cal J}_i^+$ span
$sl(M+1,{\bf R})$. The representation (\ref{slMR}) acts on functions of $M$ variables.
This representation is irreducible: all its  Casimir operators are constants.

{

If $N$ is non-negative integer, a finite-dimensional representation space occurs,
\begin{equation}
\label{PM}
     {\cal P}_{N}\ =\ \langle u_1^{p_1} u_2^{p_2} u_3^{p_3} \ldots u_M^{p_M} \vert \ 0 \le p_1+p_2+p_3+ \ldots +p_M \le N \rangle\ .
\end{equation}

In order to make the representation of (\ref{eqn20}) explicit the variables $u$
should be identified with variables $\rho$,
\[
 u_1\equiv\rho_{12}\ ,\qquad u_2\equiv\rho_{13}\ ,\ \ldots\ ,\ u_k \equiv\rho_{ij} \ ,\ \ldots\ ,\
 u_{M}\equiv\rho_{n-1,n} \ ,
\]
thus, the running index $k$ is identified with $\{ij\}$ with $i < j$ and $k=1,2,3, \ldots, M$.

\begin{theorem}
The operator $\De_{rad}$ can be rewritten in terms of the generators of the
algebra $sl(M+1,R)$\,, where $M=\frac{n(n-1)}{2}$ (or precisely speaking,
in generators of its maximal affine subalgebra in representation by the
first order differential operators).
\end{theorem}

\medskip\noindent {\bf Proof}: By direct calculation.

Explicitly, the operator (\ref{eqn20}) looks as following
\begin{equation}
\label{HRexN}
   \De_{rad}({\cal J}) \ = \ P^{(2)} (J)\ +
\end{equation}


\[
   2\,\sum_{i<j\,,\, \{ij\} = k=1}^{M+1} \bigg(\frac{m_i+m_j}{m_i m_j} \bigg) {\cal J}_{k}^0\,{\cal J}_k^-\
      +\ d \, \sum_{\{ij\} = k ,\ i<j}^{M+1}\,\bigg(\frac{m_i+m_j}{m_i m_j}\bigg)\,{\cal J}_k
\] }
where $P^{(2)} (J)$ is a quadratic polynomial in $sl(M+1,R)$ generators. \quad $\Box$

%
%
%
%
%
%

\subsection{Example 1. Three-body case: $n=3$, $d\ge 2$}

\vskip 0.7cm

We provide more details for the case $n=3$, $d\ge 2$. Equations (\ref{Jacobi}) become
\be
\label{example}
 {\bf r}_1^{(F)}=\sqrt{\frac{m_1m_2}{m_1+m_2}}\left({\bf r}_2-{\bf r}_1\right)\ ,\quad m_1{\bf r}_1+m_2{\bf r}_2+m_3{\bf r}_3=0\ ,
\ee
\[
{\bf r}_2^{(F)}=\sqrt{\frac{m_2(m_1+m_2)}{m_1+m_2+m_3}}\left({\bf r}_3-\frac{m_1{\bf r}_1}{m_1+m_2}-\frac{m_2{\bf r}_2}{m_1+m_2}\right)\ .
\]
Equations (\ref{rdef}) and (\ref{quadrelations}) are now
\[
 y_1^j=r\,O_{j,d}(\phi_\ell),\quad y_2^j=b\,O_{j,d-1}(\phi_\ell)+a\,O_{j,d}(\phi_\ell)\ ,
\]
and
\[ \sum_{j=1}^dy_j^{1}y_j^{1}=(a_1)^2,\quad  \sum_{j=1}^dy_j^{1}y_j^{2}=a_1a_2,\quad  \sum_{j=1}^dy_j^{2}y_j^{2}=(a_2)^2+(a_3)^2.\]
 A calculation yields
\be
\label{xjdera}
\pa_{y_j^{1}}\ =\  O_{j,d}\pa_{a_1}+\frac{a_3O_{j,d-1}}{a_1}\pa_{a_2}-\frac{a_2O_{j,d-1}}{a_1}\pa_{a_3}+\Omega^{(1)}_j,
\ee
\be
\label{yjdera}
 \pa_{y_j^{2}}= O_{j,d}\partial_{a_2}+O_{j,d-1}\partial_{a_3} +\Omega^{(2)}_j\ .
\ee
Although we do not have  explicit expressions for the operators $\Omega_j^{(k)}$, $k=1,2$,
we can use expressions (\ref{xjdera}), (\ref{yjdera}) to define
their actions. Indeed, applying both sides of (\ref{xjdera}) to $x_j$, $y_j$, respectively,
doing the same for (\ref{yjdera}),  summing on $j$ and using the fact that $O$ is an orthogonal matrix, we obtain the following identities:
\be
\label{xid}
 \sum_{j=1}^d \Om_j^{(x)}O_{j,d}=\frac{d-1}{r},\quad \sum_{j=1}^d \Om_j^{(x)}O_{j,d-1}=\frac{a(2-d)}{br}\ ,
\ee
\be
\label{yid}
 \sum_{j=1}^d \Omega_j^{(y)}O_{j,d}=0,\quad \sum_{j=1}^d \Omega_j^{(y)}O_{j,d-1}=\frac{d-2}{b}\ .
\ee
Thus, these sums are independent of the angular variables. From these results, using (\ref{xjdera}), (\ref{yjdera}), (\ref{xid}),(\ref{yid}), and the orthogonality of the $O$-matrices, it is straightforward to show that the Laplacian splits in the form
\[
 \De_R\ =\ \De_{\rm rad}+\Om\ ,
\]
where the operator $\Delta_{\rm rad}$ depends on the radial variables (\ref{rel-coord}) only and the operator $\Om$ annihilates any function of the radial variables alone.
We find \cite{paper1,paper2},
\be
\label{reLapl2a}
 \De_{rad}\ =\ 2\frac{m_1+m_3}{m_1m_3}\rho_{13}\,\pa_{\rho_{13}}^2+2\frac{m_1+m_2}{m_1m_2}\rho_{12}
\,\pa_{\rho_{12}}^2 + 2\frac{m_2+m_3}{m_2m_3}\rho_{23}\,\pa_{\rho_{23}}^2
\ee
\[
 +\ \frac{2}{m_1}(\rho_{13}+\rho_{12}-\rho_{23})\,\pa_{\rho_{13}\rho_{12}}
 +\ \frac{2}{m_3}(\rho_{13}+\rho_{23}-\rho_{12})\,\pa_{\rho_{13}\rho_{23}}\ +\
 \frac{2}{m_2}(\rho_{23}+\rho_{12}-\rho_{13})\,\pa_{\rho_{23}\rho_{12}}\]
 \[ +d\,\frac{m_1+m_3}{m_1m_3}\,\pa_{\rho_{13}}+d\,\frac{m_1+m_2}{m_1m_2}\pa_{\rho_{12}}+
  d\,\frac{m_2+m_3}{m_2m_3}\pa_{\rho_{23}} \ ,
\]
see (\ref{eqn20}). It is evident that this operator can be rewritten in terms of $sl(4, {\bf R})$ generators, see the representation (\ref{slMR}) at $n=3$.

The contravariant metric tensor
\[g^{\mu \nu}\ = \ \left(
\begin{array}{ccc}
 2\frac{\left(m_1+m_2\right)}{m_1\, m_2}\rho_{12} & \frac{\rho _{12}+\rho _{13}-\rho _{23}}{m_1} & \frac{\rho _{12}-\rho _{13}+\rho _{23}}{m_2} \\
 \frac{\rho _{12}+\rho _{13}-\rho _{23}}{m_1} & 2\frac{\left(m_1+m_3\right)}{m_1\, m_3}\rho_{13} & \frac{\rho _{13}+\rho _{23}-\rho _{12}}{m_3} \\
 \frac{\rho _{12}-\rho _{13}+\rho _{23}}{m_2} & \frac{\rho _{13}+\rho _{23}-\rho _{12}}{m_3} & 2\frac{\left(m_2+m_3\right)}{m_2 \, m_3}\rho_{23} \\
\end{array}
\right) \ ,
\]
in these coordinates is positive definite, does not depends on $d$ and its determinant is
\[
  D_3\ =\ \det g^{\mu \nu}\ =\ 2\,\frac{m_1+m_2+m_3}{m_1^2m_2^2m_3^2} \times
\]
\begin{equation}
\label{gmn33-rho-det-M}
 \left(m_1 m_2\rho_{12}+m_1 m_3\rho_{13}+m_2 m_3\rho_{23}\right)
                     \left(2\rho_{12}\rho_{13} + 2 \rho_{12}\rho_{23} + 2 \rho_{13}\rho_{23}-\rho_{12}^2- \rho_{13}^2 - \rho_{23}^2\right) \ .
\end{equation}
It is worth noting a remarkable factorization property of the determinant \cite{paper1,paper2},
\[
D_3\ =\ 2\frac{m_1+m_2+m_3}{m_1^2m_2^2m_3^2} \,(m_1 m_2 r_{12}^2+m_1 m_3 r_{13}^2+m_2 m_3 r_{23}^2)\ \times
\]
\[
(r_{12}+r_{13}-r_{23})(r_{12}+r_{23}-r_{13})(r_{13}+r_{23}-r_{12})(r_{12}+r_{13}+r_{23})\ =
\]
\[
   =\ 32\, \frac{m_1+m_2+m_3}{m_1^2 m_2^2 m_3^2}\ S^2_{\triangle} \ P_m \ \equiv\ c_3(m) F_1 F_2\ ,
\]
where $P_m=m_1 m_2 r_{12}^2+m_1 m_3 r_{13}^2+m_2 m_3 r_{23}^2$ - the weighted sum of squared of sides of the interaction triangle and ${S}_{\triangle}$ is their area. Hence, $D_3$ is proportional to  ${S}_{\triangle}^2\equiv {V}_3^2=F_1$, see below Conjecture 2,
and $c_3(m)=c_1 c_2\, \frac{m_1+m_2+m_3}{m_1^2 m_2^2 m_3^2}$, where
$c_k = 2^k (k!)^2$ at $k=1,2$. For the case of equal masses $m_1=m_2=m_3=1$ the factor $c_3(1)=96$,
$F_1={V}_3^2$, thus, remains unchanged, see below Conjecture 2, is square of the area of triangle,
$F_2=\sum {V}_2^2$ is the sum of all three squares of the distances between bodies, edges of the interaction triangle.

Making the gauge transformation of (\ref{reLapl2a}) with determinant (\ref{gmn33-rho-det-M}) inspired gauge factor \cite{paper2},
\begin{equation}
         \Gamma \ = \  F_1^{\frac{2-d}{4}}\,F_2^{-\frac{1}{4}}\ =\  (S^2_{\triangle})^{\frac{2-d}{4}}\,(P_m)^{-\frac{1}{4}} \ ,
\label{factor3}
\end{equation}
we find that
\begin{equation}
         \Gamma^{-1}\, {\De_R}(\rho_{ij})\,\Gamma \ =
        \  \De_{LB}(\rho) - V_{\rm eff} \ ,
\label{HLB3M}
\end{equation}
is the Laplace-Beltrami operator with the effective potential \cite{paper2},
\[
{V_{\rm eff}} \ =\ \frac{3}{8}\ \frac{(m_1+m_2+m_3)}{\left(m_1 m_2 \rho_{12}+m_1 m_3 \rho_{13}+m_2 m_3 \rho_{23}\right)}\ -
\]
\[
 \frac{(d-2)(d-4)}{2}\ \frac{\left(m_1 m_2 \rho_{12}+m_1 m_3 \rho_{13}+m_2 m_3 \rho_{23}\right)}
  { m_1 m_2 m_3\left(\rho_{12}^2+\rho_{13}^2+\rho_{23}^2 -2 \rho_{12} \rho_{13}-
                     2 \rho_{12} \rho_{23}-2 \rho_{13} \rho_{23}\right)}\ ,
\]
or, in geometrical terms,
\[
{V_{\rm eff}} \ =\ \frac{3}{8}\ \frac{(m_1+m_2+m_3)}{P_m}\ +\frac{(d-2)(d-4)}{2}\
  \frac{P_m}{m_1 m_2 m_3\,S^2_{\triangle}}
\]
or, equivalently,
\begin{equation}
{V_{\rm eff}} \ =\ \frac{3}{8}\ \frac{(m_1+m_2+m_3)}{F_2}\ +\frac{(d-2)(d-4)}{2}\
  \frac{P_m}{m_1 m_2 m_3\,F_{1}}\ ,
\label{Veff3}
\end{equation}
where the second, singular at $S^2_{\triangle}=0$ term for reasons unclear so far to present authors
is absent for $d=2,4$.
The Laplace-Beltrami operator plays a role of the kinetic energy of three-dimensional quantum particle moving in curved space in the potential (\ref{Veff3}). Seemingly, this potential looks like a three-body generalization of the centrifugal potential, which is exactly the case for two-body problem.

For $n=3$ the symmetry algebra for the reduced Hamiltonian is $so(2)$. To find it
we consider the subset consisting of the single
\[
{\tilde K}_{12}=\sum_{\ell=1}^d \left(y_{1}^{\ell}\pa_{y_{2}^{\ell}}-y_{2}^{\ell}\pa_{y_{1}^{\ell}}\right)
 \ .
\]
The $ {\tilde K}_{12}$ generates the algebra $so(2)$. (Recall that the $y_k^\ell$ are defined by (\ref{quadrelations2}).)
Using the relations (\ref{xid}), (\ref{yid}), and the orthogonality of the matrix $O$, it follows easily that the symmetry decomposes as
\[
{\tilde K}_{12}=K_{12} +\Om_{12}\ ,
\]
where $K_{12}$ depends on the radial variables $\bf a$ alone and the operator $\Omega_{12}$ annihilates any function of the radial variables alone.  In particular,
\[
K_{12}=-a_2\pa_{a_1}+\frac{(a_1)^2-(a_3)^2}{a_1}\pa_{a_2}+\frac{a_2a_3}{a_1}\pa_{a_3}\ .
\]
Solving equations (\ref{example}) for the variables $\rho_{12},\rho_{13},\rho_{23}$, we find

\bea
\rho_{12}&=&(a_1)^2\frac{m_1+m_2}{m_1m_2}\ ,\\
(m_1+m_2)\rho_{13}&=&\bigg((a_2)^2+(a_3)^2\bigg)\frac{m_1+m_2+m_3}{m_3}\nonumber \\&+&a_1^2\frac{m_2}{m_1}+2a_1a_2
\sqrt{\frac{m_2(m_1+m_2+m_3)}{m_1m_3}}\ ,\nonumber \\
(m_1+m_2)\rho_{23}&=&(a_1)^2\frac{m_1}{m_2}-2a_1a_2\sqrt{\frac{m_1(m_1+m_2+m_3)}{m_2m_3}}\nonumber\\&+&
\bigg((a_2)^2+(a_3)^2\bigg)\frac{m_1+m_2+m_3}{m_3}\ .\nonumber
\eea

Changing variables in the expression for $K_{12}$ we find $K_{12}=c\,L$, where
\[
L\ =\ -m_3\left[(m_1-m_2)\rho_{12}+(m_1+m_2)\rho_{13}-(m_1+m_2)\rho_{23}\right]\partial_{\rho_{12}}
\]
\[
+\ m_2\left[(m_1+m_3)\rho_{12}+(m_1-m_3)\rho_{13}-(m_1+m_3)\rho_{23}\right]\partial_{\rho_{13}}
\]
\be
\label{K12}
-\ m_1\left[(m_2+m_3)\rho_{12}-(m_2+m_3)\rho_{13}+(m_2-m_3)\rho_{23}\right]\partial_{\rho_{23}}\ ,
\ee
and $ c=[m_1m_2m_3(m_1+m_2+m_3)]^{-1/2}$\,.

By construction, the 3-body Hamiltonian (\ref{Hgen}) is essentially self-adjoint with respect to the volume measure $dv=\prod_{j=1,2,\, 1\le k\le d}dy_j^k$.
Moreover,
by examining the Jacobian for the change of variables (\ref{xjdera}),(\ref{yjdera}), one can verify that the transformed volume measure takes the form
\[
dv\,=\, (a_1)^{d-1}(a_3)^{d-2} da_1\ da_2\ da_3\ d\Omega \ ,
\]
where $d\Omega$ is the angular part of the measure. It follows that the  reduced Hamiltonian $-\De_{\rm rad}+V$ is essentially self-adjoint with respect to
the radial measure $dv_{\rm rad}\,=\, (a_1)^{d-1}(a_3)^{d-2} da_1\, da_2\, da_3$. Changing variables again to $\rho_{12},\rho_{13},\rho_{23}$, we obtain the normalized  radial volume measure
\be
\label{normmeasure}
dv_{\rm rad} \ =\ \left(2\rho_{12}\rho_{13}+2\rho_{12}\rho_{23}+2\rho_{13}\rho_{23}-\rho_{12}^2-\rho_{13}^2
-\rho_{23}^2\right)^{\frac{d-3}{2}}\,d\rho_{12}\,d\rho_{13}\,d\rho_{23}\ ,
\ee
which can be immediately recognized as the square of area of interaction triangle (\ref{gmn33-rho-det-M})
in degree $\frac{d-3}{2}$. It is not surprising that is of pure geometrical nature, it does not contain any mass dependence.
Although the radial reduced Hamiltonian is essentially self-adjoint it is not in the form of a Laplace-Beltrami operator plus potential. For this a further gauge transformation (\ref{HLB3M}) with (\ref{factor3}) is needed, see \cite{paper2}.

\subsection{Example 2: More detail on the equal mass case $n=4$, $d\ge 3$}

Here $\Delta_{\rm rad}$ is given by (\ref{eqn20}) for $n=4$, $m_j=1$. It is evident that the determinant
of contravariant metric $g^{\mu\nu}$ is polynomial in $\rho$'s variables. It can be factorized to the product of two polynomials
\be
\label{det4}
\det g^{\mu\nu}\ =\ c_4(m=1)\, F_1\,F_2\ ,
\ee
where
\[
   c_4(m=1)\ =\ 36864\ ,\quad
F_1\ =\ {V}_4^2\ ,\quad
F_2\ =\ {\tilde V}_2^2\ {\tilde V}_3^2\ -\ 36\,{\tilde V}_1^2\ {\tilde V}_4^2\ ,
\]
and
\begin{itemize}
  \item ${\tilde V}_4^2={V}^2_4 \geq 0$ is the square of the volume  of the tetrahedron of interaction.
  \item ${\tilde V}_3^2 = \sum^4 {V}_3^2 \geq 0 $ is the sum of all four squares of the areas of the adjacent to three vertices, interaction triangles of tetrahedron.
  \item ${\tilde V}_2^2 = \sum^6 {V}_2^2 \geq 0$ is the sum of all six squares of the distances between
         bodies, adjacent to two vertex edges of the interaction tetrahedron.
  \item By definition ${\tilde V}_1^2 \equiv \sum {V}_1^2 \equiv 1$.
  \item It can be proved that ${\tilde V}_2^2\ {\tilde V}_3^2\ \geq \ 36\,{\tilde V}_1^2\ {\tilde V}_4^2$.
\end{itemize}
Lower index marks the number of vertices of the face considered.
Hence, both $F_{1,2}$ are of geometrical nature. They define the boundary of the configuration space,
$F_1=0$, where the determinant  vanishes.
Now, one can find the gauge factor $\Gamma$ such that the operator $\Delta_{\rm rad}$ takes
the form of the Schr\"odinger operator,
\begin{equation}
\label{DLB4}
     \Gamma^{-1}\,\Delta_{\rm rad}\, \Gamma\ =\ {\Delta_{LB}} - { V_{\rm eff}}\ ,
\end{equation}
where $\De_{LB}$ is the Laplace-Beltrami operator with contravariant metric $g^{\mu\nu}$.
We obtain
\begin{equation}
       \Gamma\ =\ (F_1 F_2)^{-1/4}({\tilde V}_4^2)^{1-d/4}\ =\ F_1^{\frac{3 - d}{4}}\,F_2^{-\frac{1}{4}}\ ,
\label{factor4}
\end{equation}
c.f. (\ref{factor3}), therefore $\Gamma$ is made from degrees of factors appeared in determinant (\ref{det4}), and the effective potential is
\[
V_{\rm eff}\ =\ \frac{3\,({\tilde V}_2^2)^2\
+\ 112\, {\tilde V}_3^2}{32\,({\tilde V}_2^2\, {\tilde V}_3^2\ -\ 36\,{\tilde V}_4^2\ )}\ +\
\frac{(d-3)(d-5)\,{\tilde V}_3^2}{72\,{\tilde V}_4^2}\ ,
\]
or, equivalently,
\begin{equation}
    V_{\rm eff}\ =\  \frac{3({\tilde V}_2^2)^2\, +\, 112\, {\tilde V}_3^2}{32\,F_2}\ +\
                      \frac{(d-3)(d-5)\,{\tilde V}_3^2}{72 F_1}
\label{Veff4}
\end{equation}
c.f. (\ref{Veff3}),
where the first term is $d$-independent while the second term vanishes for $d=3,5$.
It contains the factors $F_{1,2}$ as denominators, see (\ref{det4}). Hence, the effective potential becomes singular at boundary of the configuration space. Overall, the effective potential is of geometrical nature made out of the volumes of faces of 4-vertex polytope.

For $n=4$ and equal masses the reduced radial Laplacian (\ref{eqn20}) admits a 3-dimensional symmetry algebra with elements proportional to
\[
  L(a,b,c)\ =
\]
\[
   \left (\rho_{13}\,a+\rho_{14}\,b-\rho_{23}\,a-\rho_{24}\,b\right)\pa_{\rho_{12}}+\left(-\rho_{12}
   (\frac32\, a+\frac32\,b+c)+\rho_{14}(\frac32\, a+\frac72\,b+3c)+\right.
\]
\[
   \left. \rho_{23}
   (\frac32\, a+\frac32\, b+c)-\rho_{34}(\frac32\,a+\frac72\,b+3c)\right)\pa_{\rho_{13}}+\left(\rho_{12}(\frac12\, a-\frac12\, b-c)\right.
\]
\[
   +\rho_{13} (-\frac12\, a +\frac12\, b+c)+ \left.
   \rho_{24}(\frac52\, b+\frac32\, a+3c)-\rho_{34}(\frac52\, b+\frac32\, a+3c)\right)\pa_{\rho_{23}}
\]
\[
   +\bigg(c \rho_{12}-\rho_{13}(a+3b+3c)-\rho_{24}\,c +
   \rho_{34}(a+3b+3c)\bigg)
     \pa_{\rho_{14}}+\left(\rho_{12}(2b+c+a)-\right.
\]
\[
  \left.\rho_{14}(2b+c+a)-\rho_{23}(2a+3b+3c)+\rho_{34}(2a+3b+3c)\right)\pa_{\rho_{24}}+
\]
\[
  \left(\rho_{13}(\frac12\, a
      +\frac52\, b+2c)-\rho_{14}(\frac12\, a +\frac52\, b+2c)+
    \rho_{23}(\frac32 \, a+\frac32\, b+2c)\right.
\]
\begin{equation}
\label{Lparam}
\left. -\rho_{24}(\frac32\, a+\frac32\, b+2c)\right)\pa_{ \rho_{34}}\ ,
\end{equation}
where $a,b,c$ are arbitrary parameters. The symmetry $L(2,-2,1)$ is convenient for separating a variable. A suitable basis is $\{J_1,J_2,J_3\}$, where
\[
  J_1\ =\ L\,(\frac{2}{\sqrt{11}},0,0)\ ,\
  J_2\ =\ L\,\left(\frac{\sqrt{3}}{12}(-1+\frac{17}{\sqrt{11}})\ ,
  -\frac{\sqrt{3}}{12}(5+\sqrt{11}),
  \frac{\sqrt{3}}{2}\right)\ ,
\]
\[
 J_3\ =\ L\,\left(\frac{\sqrt{3}}{12}(1+\frac{17}{\sqrt{11}}),\frac{\sqrt{3}}{12}(-5+\sqrt{11}),
\frac{\sqrt{3}}{2}\right)\ .
\]
This basis satisfies the commutation relations
\[
[J_1,J_2]=J_3,\ [J_2,J_3]=J_1,\ [J_3,J_1]=J_2\ ,
\]
so the symmetry algebra is isomorphic to $so(3,R)$.The gauge factor is invariant under all these symmetries so they remain unchanged  as symmetries  for the Laplace-Beltrami operator.

\section{The determinant.\\ The Laplace-Beltrami operator (conjectures)}

In general, the operator $\De_{rad}$ (\ref{eqn20}) can be written as
\begin{equation}
\label{detgmn}
  \De_{rad}\ =\ g^{\mu \nu} \pa_{\mu} \pa_{\nu}\ +\  d\, \sum_{i<j}^n\,\bigg(\frac{m_i+m_j}{m_i m_j}\bigg)\pa_{\rho_{ij}} ,
\end{equation}
c.f. (\ref{gmunu}), where $g^{\mu \nu}$ is matrix made out of coefficients in front of the second derivatives and  the remaining terms involve the  first derivatives.
We will make sense of $g^{\mu \nu}$ as the contravariant metric tensor, see below.

The matrix $g^{\mu \nu}$ does not depends on $d$ and is of the size $M \times M$, where $M=\frac{n(n-1)}{2}$. Vector $b^{\mu}$ is coordinate-independent, it is proportional to $d$ and contains two-body reduced masses.

\begin{conjecture}
The determinant $\det g^{\mu \nu}$ is a homogeneous polynomial in $M$ variables $\rho_{ij}$ of degree $M=\frac{n(n-1)}{2}$, it  factors as  the product of two homogeneous polynomials in $\rho_{ij}$
\be
\label{det-n}
\det g^{\mu \nu}\ =\ c_n(m)\, F_1\,F_2\ ,
\ee
where $c_n(m)$ is   mass-dependent,
\[
  c_n(m)\ =\ c_1 c_2 \ldots c_{n-1}\ \frac{m_1+m_2+\ldots+m_n}{(m_1 m_2 \ldots m_n)^2}\ ,\quad c_k=2^k (k!)^2\ ,
\]
and
\[
      F_1\ =\ {\tilde V}_n^2\ ,
\]
is the square of the volume of the $n$-vertex polytope of interaction  given by the Cayley-Menger determinant\cite{CayleyMenger}, hence, a homogeneous polynomial in $\rho$ of degree $(n-1)$, and $F_2$ is a homogeneous polynomial in $\rho$ of degree $\frac{(n-1)(n-2)}{2}$  written as,
\[
F_2\ =\ Polynomial({\tilde V}^2;m)\ \equiv
\]

\[
     P\bigg[{\tilde V}^2_n\,, {\tilde V}^2_{n-1}, {\tilde V}^2_{n-2} \ldots
    {\tilde V}^2_{2}\,, {\tilde V}^2_{1} \bigg]\ ,
\]
(where by definition ${\tilde V}^2_{1}=1$, and ${\tilde V}^2_n=V^2_n$ is the square of the volume of a $n$-vertex polytope), is a polynomial in weighted sums over squares of the volumes of faces of given dimension, or equivalently, of given number $(n-k)$ of vertices, ${\tilde V}^2_{n-k} \geq 0$ with mass $m$-dependent coefficients. Hence, $F_2$ depends effectively on $(n-1)$ polynomial variables $\{\tilde V\}$, $F_2 \geq 0$.
\end{conjecture}

It was checked that this conjecture is valid for $n=2,3,4$ and arbitrary masses, and $n=5, 6$ for equal masses.
\begin{conjecture}
There exists the gauge factor
\begin{equation}
       \Gamma\ =\ F_1^{\frac{n - 1 - d}{4}}\,F_2^{-\frac{1}{4}}\ ,
\label{factor-n}
\end{equation}
with $F_{1,2}$ from Conjecture 2,
such that the operator $\De_{\rm rad}$ takes the form of the Schr\"odinger
operator,
\begin{equation}
\label{DLB-n}
     \Gamma^{-1}\,\De_{\rm rad}\, \Gamma\ =\ {\Delta_{LB}} - { V_{\rm
eff}}\ ,
\end{equation}
where $\De_{LB}$ is the Laplace-Beltrami operator with contravariant
metric $(-g^{ij})$ and
$V_{\rm eff}$ is the effective potential which is a superposition of two
ratios containing $F_{1,2}$ as denominators,
\[
V_{\rm eff}\ =\ \frac{P({\tilde V}_1^2, \ldots\ , {\tilde V}_{n}^2)}{F_2}\ +\
\frac{(d-n+1)(d-n-1)\,Q({\tilde V}_1^2, \ldots\ ,{\tilde V}_{n-1}^2)}{F_1}\ ,
\]
where $P(\rho)$ and $Q(\rho)$ are homogeneous polynomials of degrees
$\frac{n(n-3)}{2}$ and $(n-2)$, respectively.
\end{conjecture}

It was checked that this conjecture is valid for $n=2,3,4$. In particular,
for $n=2$ the effective potential is
\[
    V_{\rm eff}\ =\ \frac{(d-1)(d-3)\,(m_1+m_2)}{8\, m_1\, m_2\,
\rho_{12}}\ ,
\]
while for $n=3$ it is given (\ref{Veff3}).

 Assuming that Conjectures 2 and 3 are correct, it follows easily that the radial reduced Hamiltonian $(-\De_{\rm rad} + V)$ of the  $n$-body problem is essentially self-adjoint with respect to the normalized radial measure (\ref{vrad-A}) written in $\rho$-variables of the form
\begin{equation}
\label{vrad-rho}
      dv_{\rm rad}\ =\ (V_n^2)^{\frac{d-n}{2} } \prod_1^{\frac{n(n-1)}{2}}\, d \rho_{ij}\ ,
\end{equation}
where $V_n^2$ is square of the volume of the $n$-vertex polytope.

\section{Towards classical systems: De-quantization}

We show that many of the results derived above hold for related classical systems that we consider as de-quantization of the quantum systems, i.e., we make the replacement
\[
       -i\pa_{{\bf r}_j}\to p_{{\bf r}_j}\ ,
\]
thus the quantum momentum is replaced by classical one.
Then instead of (\ref{Dflat}), with the center of mass coordinates split off, we have
\begin{equation}
\label{classicalflat}
       {\cal H}_r^{(d(n-1))}\ =\ p_{{\bf r}_i^{(F)}} p_{{\bf r}_i^{(F)}}.
\end{equation}

We introduce radial and angular variables as before. Equation (\ref{xjderb}) becomes
\begin{equation}
\label{xjmom}
p_{{\bf x}_j^k}=\sum_{\ell=1}^N S^k_j({\bf a},O,\ell)p_{a_\ell}+\Om_j^{(k)}\ ,
\end{equation}
where \[
S_j^k({\bf a},O,\ell)=\sum_{s=1}^NT({\bf a},\ell,k)_s\, O_{j,d-s+1}\ ,
\]
the $T({\bf a},\ell,k)_s$ are simple rational functions of the radial variables alone,
\be
\label{Omegaopc}
 \Omega^{(k)}_j=\sum_{\ell=1}^{\frac{(n-1)(2d-n)}{2}}\frac{\partial \phi_\ell}{\partial {\bf x}_j^k}p_{\phi_\ell},
\ee
and the $\phi_\ell$ are the angular variables.
Using this result and the orthogonality of the $O$ matrix we find
\begin{equation}
\label{classicalsplit}
  {\cal H}_r^{(d(n-1))} =p_{{\bf r}_i^{(F)}} p_{{\bf r}_i^{(F)}}={\cal H}_{\rm rad} +\Omega,
\end{equation}
2nd order homogeneous in the momenta, where ${\cal H}_{\rm rad} $ depends on the radial variables and momenta only and every term in $\Omega$ is either a product of 2 angular momenta or a product of a radial momentum and an angular momentum.

\begin{equation}
\label{HRN1}
  {\cal H}_r^{(d(n-1))}=\  2\,\sum_{i<j}^n \bigg(\frac{m_i+m_j}{m_i m_j} \bigg) \rho_{ij} p^2_{\rho_{ij}}\ +\ 2\,\sum_{i \neq j,i\neq k, j< k}^n\,\frac{1}{m_i}\,(\rho_{ij} + \rho_{ik} - \rho_{jk})p_{\rho_{ij}}p_{\rho_{ik}}\ +\ \Omega  \ ,
\end{equation}

Now we introduce alternate  radial coordinates
\[
\rho_{\ell k}=\sum_{s=1}^d(x_{\ell,s}-x_{k,s})^2,\quad 1\le \ell<k\le n\ .
\]
Thus
\[
p_{x_{\ell,s}}=2\sum_{h\ne l}(x_{\ell,s}-x_{h,s})p_{ \rho_{\ell h}}+\cdots
\]
where the non-explicit terms are  momenta in the angular and center of mass variables. From this we can compute the coefficient of $p_{ \rho_{\ell k}}^2$
in $\sum_{i=1}^n\frac{1}{2m_i} p_i^{(d)}$ for $\ell<k$. It is
\[
  2\bigg(\frac{m_\ell+m_k}{m_\ell m_k}\bigg)\sum_{s=1}^d(x_{\ell,s}-x_{k,s})^2=4\rho_{\ell k}\ .
\]
Similarly the coefficient of  $p_{ \rho_{\ell k}}p_{ \rho_{\ell {k'}}}$ for $k< k'$ is
\[
  \frac{ 4}{m_\ell}\sum_{s=1}^d(x_{\ell,s}-x_{k,s})(x_{\ell,s}-x_{{k'},s})\
  =\ \frac{2}{m_\ell}(\rho_{\ell {k}}+\rho_{\ell {k'}}-\rho_{k {k'}})\ .
\]
The coefficient of $p_{ \rho_{\ell k}}p_{ \rho_{{\ell'} {k'}}}$ for $k,{k'},\ell,{\ell'}$
pairwise distinct is $0$. Thus we have determined all of the terms in ${\cal H}_{\rm rad}$.

The Hamilton-Jacobi equation is
\begin{equation}
\label{HJa}
 {\cal H}_r ^{(d(n-1))}+V(\rho_{ij})\ =\ E\ ,
\end{equation}
where
\[
   p_{\rho_{ij}}=\frac{\pa W}{\pa \rho_{ij}},\  p_{\phi_k}\
   =\ \frac{\partial W}{\pa{\phi_k}}\ ,
\]
for $W=W(\rho_{ij},\phi_k)$. Now we restrict the equation by requiring  that $W$ is a function
of the radial variables alone: $W=f(\rho_{ij})$. The restricted Hamilton-Jacobi equation is

\begin{equation}
\label{HJb}
   2\,\sum_{i<j}^n \bigg(\frac{m_i+m_j}{m_i m_j}\bigg) \rho_{ij} p^2_{\rho_{ij}}  +
 \,\sum_{i \neq j,i\neq k, j< k}^n\,\frac{2}{m_i}(\rho_{ij} + \rho_{ik} - \rho_{jk})p_{\rho_{ij}}p_{\rho_{ik}} +V(\rho_{ij})\ =\ E\ ,
\end{equation}
which can be considered as describing a  classical top with variable tensor of inertia
in an external potential, \cite{Arnold}.
This is one interpretation of  de-quantization. The second interpretation is to apply the same procedure to the quantum equations after the gauge transformation to put the Hamiltonian in self-adjoint form, so that the effective potential has been introduced. Then the de-quantized Hamilton-Jacobi equation contains the effective potential.

\section{Conclusions}

In this paper, studying the quantum $n$ body problem in $d$-dimensional space $d > n-2$, we have made the change of variables
from individual Cartesian coordinates $\{ \vec{r} \}$ to centre-of-mass vector coordinate $\vec{R}_{0}$, mutual distances between bodies $\{ r_{ij} \}$ and angles ${\Theta}$,
\[
    (\vec{r_1}, \vec{r_2}, \ldots , \vec{r_n}) \rar \bigg(\vec{R}_{0}, \{ r_{ij} \}, \Theta \bigg)\ .
\]
We assumed that the total relative angular momentum vanishes.
As a result the kinetic energy (the flat diagonal Laplace operator) decomposes naturally
into the sum of three operators
\[
 \sum_{i=1}^n\frac{1}{2m_i} \De_i^{(d)}\ =\ \De_{{\bf R}_{0}} + \De_{\rm rad} + \Om \ ,
\]
where $\De_{{\bf R}_{0}}$ is the center of mass Laplacian,
the operator $\De_{\rm rad}$ depends on the mutual distances $\equiv$ radial variables, ${\rho_{ij}=r_{ij}^2}$, alone, and $\Om$ annihilates any function of the radial variables alone.
Moreover, the operator $\De_{\rm rad}(\rho)$ is stable,  independent of how angular variables are introduced. It contains very simple mass dependence;
it is algebraic, see (\ref{eqn20}), and $sl(M+1, R)$-Lie-algebraic, where $M=\frac{n(n-1)}{2}$,
see (\ref{HRexN}).  It must be emphasized that unlike Jacobi coordinates
(the Jacobi relative distances) the relative (mutual, radial) distances $r_{ij}$ do {\bf not}
contain mass dependencies.
Seemingly, $\De_{\rm rad}(\rho)$ is self-adjoint for any number of bodies $n$ - it was checked
constructively for $n=2,3,4$ - for arbitrary masses.

If we consider the family of angle-independent eigenfunctions, the above-mentioned change of variables implies that the original $n$-body spectral problem,
\[
    {\cal H} \Psi\ =\ E \Psi\ ,
\]
is reduced to a {\it restricted} spectral problem,
\begin{equation}
\label{restricted}
   \bigg(-\De_{\rm rad}(\rho) + V(\rho)\bigg) \Psi\ =\ E \Psi\ .
\end{equation}
The restricted spectral problem looks much simpler than the original one, it depends on $M=\frac{n(n-1)}{2}$ variables solely.
Certainly, the ground state function, if it exists, should be an eigenfunction of the restricted spectral problem as was predicted by Ter-Martirosyan \cite{Ter}. It implies that the spectra of
angle-independent eigenfunctions of the original problem coincides with the spectra of the restricted problem.

\begin{theorem}
Suppose that the  Hamiltonian operator ${\cal H}_r$ of the $n$-body spectral problem (\ref{Hrel}) is essentially self-adjoint, with nondegenerate ground state of energy $E_0$, and ground state eigenfunction $\Psi_0$.  Then $\Psi_0$ depends on relative distances (radial coordinates) $\{r_{ij}\}$ alone and is also the nondegenerate ground state for the restricted spectral problem (\ref{redrad}).
\end{theorem}

\medskip\noindent {\bf Proof}: The original Hamiltonian (\ref{Hgen}) admits the Euclidean group $E(d,R)$ as a symmetry group and the system  (\ref{Hrel}) is translation invariant and admits the orthogonal group $SO(d,R)$  as a compact symmetry group. Thus the space of bound states of the Hamiltonian ${\cal H}_r$ decomposes into a direct sum of subspaces, each subspace irreducible under the action of $SO(d,R)$ and corresponding to a specific eigenvalue. Since the ground state is nondegenerate, it must correspond to a 1-dimensional representation space, i.e., it must be invariant under rotations. Since the functions $\{ \rho_{ij}\}$ form an integrity basis for translation invariant and rotation invariant functions in the variables $\{ {\bf r}_j\}$, $\Psi_0$ must be a function of these variables alone, hence an eigenfunction of the operator (\ref{redrad}) with eigenvalue $E_0$. Since all eigenfunctions of the restricted problem are also eigenfunctions of the original problem (\ref{Hrel}), $\Psi_0$ must also be the nondegenerate ground state eigenfunction for the restricted system. $\Box$

\medskip

The operator of the restricted spectral problem (\ref{restricted}) is, in fact, the Hamiltonian of
a $\frac{n(n-1)}{2}$-dimensional (degenerate) Euler-Arnold quantum top on the algebra $sl(\frac{n^2-n+2}{2}, R)$ in a constant magnetic field with external potential $V(\rho)$, see (\ref{HRexN}), c.f. \cite{Turbiner:1988,RT:1995,Brink:1997,Turbiner:2016} with constraints that all Casimir operators of $sl(M+1,R)$ algebra are constants. It has to be emphasized that similar quantum tops in constant magnetic field occur in $A_M, BC_M, D_M$ Calogero-Moser-Sutherland models at $d=1$, see e.g. \cite{RT:1995,Brink:1997}.

If our {\it Conjecture 3} is valid in its full generality, there exists a gauge factor $\Gamma$ such that the operator of the restricted spectral problem (\ref{restricted}) is gauge-equivalent to the Hamiltonian of
$\frac{n(n-1)}{2}$-dimensional quantum particle in curved space in external potential,
\[
   \Gamma^{-1} \bigg( -\De_{\rm rad}(\rho) + V(\rho)\bigg) \Gamma \ =\ -\De_g + V_{\rm eff}(\rho) + V(\rho)
   \ \equiv \ {\cal H}_{\rm rad}\ .
\]
Here $\De_g$ is the Laplace-Beltrami operator with contravariant metric $g^{\mu \nu}$ given by matrix of the coefficients in front of second derivatives in $\De_{\rm rad}(\rho)$ (\ref{eqn20}), and $V_{\rm eff}(\rho)$ is the effective potential which emerged as a result of gauge rotation. The boundary of the configuration space for ${\cal H}_{\rm rad}$ is defined by the condition $\det g^{\mu \nu}=0$.

The (Lie)-algebraic form of the operator $\De_{\rm rad}(\rho)$ suggests a  direction for finding the exact solutions of both restricted and original spectral problems. In particular, adding to $\De_{\rm rad}(\rho)$ the terms linear in derivatives, $A_{ij} \rho_{ij} \pa_{ij}$ (which is the Euler operator made from the elements of the Cartan subalgebra of $sl(\frac{n^2-n+2}{2}, R)$), and then gauging them away with factor $\sim \exp ( - {\tilde A}_{ij} \rho_{ij})$ leads to the harmonic oscillator potential in the space of relative distances,
\[
     V^{(ex)}\ =\ \om_{ij} \rho_{ij}\ ,
\]
which is an exactly-solvable potential for the restricted problem and perhaps, quasi-exactly-solvable for the original problem. In general, the question about (quasi)-exact solutions needs to be investigated separately; it will be done elsewhere.

The operator $\De_{\rm rad}(\rho)$ admits a simple limit to the atomic (say, $m_1 \to \infty$) and molecular
(say, $m_{1,\ldots,p} \to \infty$) situations. In the atomic case, for the operator $\De_{\rm rad}(\rho)$ (\ref{eqn20}) all second order cross derivatives $\pa_{\rho_{1j}}\pa_{\rho_{1k}}$ disappear, while other terms remain. The number of variables in this case remains unchanged. In the molecular case, not only cross derivatives $\pa_{\rho_{qj}}\pa_{\rho_{qk}}, q=1,\ldots,p$ but also the derivatives w.r.t. $\rho_{ij},\ 1 \leq i<j \leq p$ vanish. In total, the operator $\De_{\rm rad}(\rho)$ depends on $M - \frac{p(p-1)}{2}$ variables. Other variables which may appear in the potential $V(\rho)$ are external parameters. This corresponds to the so-called Bohr-Oppenheimer approximation (of zero order) in molecular physics.

The operator $\De_{\rm rad}(\rho)$ was derived for $d > (n-2)$. A natural question is how this operator looks like for $d = 1,2,3,\ldots, (n-2)$. At $d=1$ this operator was found in \cite{RT:1995,Brink:1997} for arbitrary $n$ and equal masses, $m_i=1$ while for the three-body case, $n=3$, in our previous paper \cite{paper2}.
It is evident that for the case $d=n-2$ the constraint $V^2_n=0$ - the square of the volume of $n$-vertex polytope of interaction vanishes - should be imposed. It will lead to the explicit form of $\De_{\rm rad}(\rho)$. The question of finding $\De_{\rm rad}(\rho)$ for $n>3$ and physical dimensions $d=2,3$ remains open.

\section*{Acknowledgments}

W.M. was partially supported by a grant from the Simons Foundation (\# 412351 to Willard Miller, Jr.).
A.V.T. is thankful to University of Minnesota, USA for kind hospitality extended to him where this work was initiated and continued during several visits, Stony Brook University, USA for kind hospitality during several visits and also to the Simons Center for Geometry and Physics, Stony Brook, USA, where this work was completed.
A.V.T. is supported in part by the PAPIIT grant {\bf IN108815}.
M.A.E.R. is grateful to ICN UNAM, Mexico and University of Minnesota, USA for the kind hospitality during his several visits, where a part of the research was done, he was supported in part by DGAPA grant
{\bf IN108815} (Mexico). The authors, all or some of them, have benefited from  discussions with A.~Abanov, V.~Korepin, R.~Moeckel, R.~Montgomery, P.~Olver, V.~Reiner, E.~Shuryak, L.~Takhtajan and D.~Zeilberger. A.V.T. thanks S.P.~Novikov for important remarks and encouragement.

Symbolic calculations were made using MAPLE-16 and MATHEMATICA-11.

\section*{Appendix: Determinant (concrete examples)}

Here we present the results of concrete calculations of the determinant of matrix $g^{\mu\nu}$, see (\ref{eqn20}), (\ref{detgmn}). For the cases $n=2,3,4$ all masses are assumed arbitrary, $m_i > 0$, while for the cases $n=5,6$ due to serious technical difficulties we assumed all masses equal, $m_i=1$.

$\bullet$\ For the two-body case, $n=2$, at $d \geq 1$ the matrix $g^{\mu\nu}$ is of the size $1 \times 1$ (single $\rho$ coordinate),
\[
g^{\mu\nu} \ = \ 2\ \bigg(\frac{m_1 + m_2}{{m_1\,m_2}}\ \rho_{12}\bigg) \ .
\]
Thus, the factors $c_1=2$, $F_1=\rho_{12}$ and $F_2=1$ explicitly occur, see Conjecture 2 and (\ref{det-n}), and
\[
D_2 \ = \ 2\, \frac{m_1 + m_2}{(m_1\,m_2)^2}\ V^2_{2} \ F_2 \ \ ,
\]
where
\[
  2\, \frac{m_1 + m_2}{(m_1\,m_2)^2}\ =\ c_2(m)\ ,\ V^2_{2} \ = \ \rho_{12}\ ,\ F_2 \ = \  m_1\,m_2 \ .
\]

$\bullet$\ For the three body case, $n=3$, at $d \geq 2$ the matrix $g^{\mu\nu}$ is of the size $3 \times 3$ (three $\rho$ coordinates) and calculations for the determinant give,
\[
D_3 \ = \ 32\ \frac{m_1 + m_2 + m_3}{(m_1\,m_2\,m_3)^2}\ V^2_{3}\ F_2 \ =\ c_3(m) F_1 F_2\ ,
\]
where $V^2_{3}=F_1$ is the square of the area of interaction triangle and
\[
F_2 \ = \ m_2\,m_3 \rho_{23} \ +  \ m_1\,m_3\, \rho_{13} \ + \ m_1\,m_2 \,\rho_{12} \ .
\]
c.f. (\ref{det-n}). For $m_1=m_2=m_3=1$ the determinant simplifies
\[
D_3(m=1)\ = \ 96\, V_3^2\, \tilde V_2^2 \ ,
\]
thus, the factor $F_2$ becomes the sum of squares of edges of the interaction triangle,
$\rho_{23}\ +\ \rho_{13}\ +\ \rho_{12}$.

$\bullet$\ For the four-body case, $n=4$, at $d \geq 3$ the matrix $g^{\mu\nu}$ is of the size $6 \times 6$
(six $\rho$ coordinates), its determinant
\[
D_4 \ = \ 9216\ \frac{m_1 + m_2+m_3+m_4}{(m_1\,m_2\,m_3\,m_4)^2}\ F_1\ F_2 \ =\ c_4(m) F_1 F_2\ ,
\]
c.f. (\ref{det-n}), where $F_1\,=\, V_4^2$ is square of the volume of tetrahedron
\[
     V_4^2 \ = \ \frac{1}{144} \times
\]
\[
   \bigg(\rho_{14} \rho_{23} \rho_{12}+\rho_{13} \rho_{24} \rho_{12}
  -\rho_{14} \rho_{24} \rho_{12}+\rho_{13} \rho_{34} \rho_{12}+\rho_{14} \rho_{34} \rho_{12}
  +\rho_{23} \rho_{34} \rho_{12} +\rho_{24} \rho_{34} \rho_{12}
  -\rho_{12} \rho_{13} \rho_{23}
\]
\[
  + \rho_{13} \rho_{14} \rho_{23} + \rho_{13} \rho_{14} \rho_{24}
  + \rho_{13} \rho_{23} \rho_{24} + \rho_{14} \rho_{23} \rho_{24}
  + \rho_{14} \rho_{23} \rho_{34} + \rho_{13} \rho_{24} \rho_{34}
  - \rho_{13} \rho_{14} \rho_{34} - \rho_{23} \rho_{24} \rho_{34}
\]
\begin{equation}
\label{F1-4}
   -\rho_{12}^2 \rho_{34} - \rho_{13}^2 \rho_{24} - \rho_{14}^2 \rho_{23} - \rho_{12} \rho_{34}^2
    -\rho_{13} \rho_{24}^2 - \rho_{14} \rho_{23}^2
    \bigg)\ ,
\end{equation}
and
\[
   F_2 \ = \ {\tilde V}^2_{3} \,{\tilde V}^2_{2}\ -\ 9\,(m_1 + m_2+m_3+m_4)\,V_4^2\ ,
\]
with
\[
\hskip -.9cm    {\tilde V}^2_{3} = \bigg[
     \frac{1}{m_1} S^2(\rho_{23},\,\rho_{24},\,\rho_{34})\,
   +\,\frac{1}{m_2} S^2(\rho_{13},\,\rho_{14},\,\rho_{34})\,
   +\,\frac{1}{m_3} S^2(\rho_{12},\,\rho_{14},\,\rho_{24})\,
   +\, \frac{1}{m_4} S^2(\rho_{12},\,\rho_{13},\,\rho_{23})\bigg]\ ,
\]
here $S^2(a^2,\,b^2,\,c^2)$ stands for the square of the area of the triangle with sides $a,b,c$ and,
\[
\tilde V^2_{2} \ = \ \bigg[{m_1\,m_2}\, \rho_{12}\ +\ {m_1\,m_3}\, \rho_{13}\ + \ {m_1\,m_4} \,\rho_{14}\ +\ {m_2\,m_3}\, \rho_{23}\ + \ {m_2\,m_4} \,\rho_{24}\ + \ {m_3\,m_4}\, \rho_{34} \bigg]\ .
\]

 For $m_1=m_2=m_3=m_4=1$ the determinant simplifies
\[
D_4(m=1)\ = \ 36864\, V_4^2\,({\tilde V}_3^2 {\tilde V}_2^2\ -\ 36 V_4^2) \ ,
\]
thus, the factor $F_2$ becomes a simple polynomial in the square of the volume of the tetrahedron,
the sum of squares of areas of interaction triangles and the sum of squares of edges of the
interaction tetrahedron.  If $V_4^2=0$, the factor $F_2 \sim D_3(m=1)$.

\vskip 0.6cm

$\bullet$\ For the five-body case, $n=5$, and $d \geq 4$  the matrix $g^{\mu\nu}$ is of size $10 \times 10$ (ten $\rho$ coordinates),
{\small
\[
 g^{\mu\,\nu}=\left(
\begin{array}{cccccccccc}
 4 \,\rho _{12} & f_{123} & f_{124} & f_{125} & f_{213} & f_{214} & f_{215} & 0 & 0 & 0 \\
 f_{123}         & 4 \,\rho _{13} &  f_{134} & f_{135} & f_{312} & 0 & 0 & f_{314} & f_{315} & 0 \\
 f_{124}         & f_{134} & 4\, \rho _{14} & f_{145} & 0 & f_{412} & 0 & f_{413} & 0 & f_{415} \\
 f_{125}         & f_{135} & f_{145} & 4 \,\rho _{15} & 0 & 0 & f_{512} & 0 & f_{531} & f_{541} \\
 f_{213}         & f_{312} & 0 & 0 & 4 \rho _{23} & f_{234} & f_{235} & f_{324} & f_{325} & 0 \\
 f_{214}         & 0 & f_{412} & 0 & f_{234} & 4\, \rho _{24} & f_{245} & f_{432} & 0 & f_{425} \\
 f_{215}         & 0 & 0 & f_{512} & f_{235} & f_{245} & 4 \,\rho _{25} & 0 & f_{532} & f_{542} \\
 0               & f_{314} & f_{413} & 0 & f_{324} & f_{432} & 0 & 4 \,\rho _{34} & f_{345} & f_{435} \\
 0               & f_{315} & 0 & f_{531} & f_{325} & 0 & f_{532} & f_{345} & 4 \,\rho _{35} & f_{543} \\
 0               & 0 & f_{415} & f_{541} & 0 & f_{425} & f_{542} & f_{435} & f_{543} & 4\, \rho _{45} \\
\end{array}
\right)
\]
}
where $f_{ijk}\equiv -\rho _{jk}+\rho _{ij}+\rho _{ik}$\,. By direct calculation using MATHEMATICA-11 its determinant can be expressed in the rather simple form,
\[
D_5(m=1)\ = \ 424673280\, V_5^2\, \bigg[-\,4 \,V_5^2\, {\bigg( \tilde V_2^2 \bigg)}^2\, +\, \tilde V_4^2\,\tilde V_3^2\,\tilde V_2^2\,-\,45\,{\bigg(\tilde V_4^2 \bigg)}^2\bigg]\ ,
\]
where $V_5^2$ is square of the volume of 5-vertex polytope given by Cayley-Menger determinant. Hence, the factors in (\ref{det-n}) can be identified as
\[
   c_5(m=1)\ =\ 424673280\ ,\  F_1\ =\ V_5^2\ ,
\]
and $F_2$ is given by the expression in square brackets,
\[
     F_2\ =\ -4 \,V_5^2\, {\bigg( \tilde V_2^2 \bigg)}^2\, +\,\tilde V_4^2\,\tilde V_3^2\,\tilde V_2^2\,-\,45\,{\bigg(\tilde V_4^2 \bigg)}^2 \geq 0\ .
\]
Thus, both factors $F_{1,2}$ are of the geometrical nature.
Note that the determinant which is a polynomial in $\rho$ with ${\deg D}_5 = 10$ is factorized
to the product of polynomials of ${\deg F}_1 = 4$ and ${\deg F}_2 = 6$.

\vskip 0.6cm

$\bullet$\ For six-body case, $n=6$, at $d \geq 5$ the matrix $g^{\mu\nu}$ is of the size $15 \times 15$ \\ (fifteen $\rho$ coordinates). After cumbersome calculation using MATHEMATICA-11 its determinant takes the unexpectedly simple factorized form,

\begin{equation}
\label{}
D_6\ =\ c_{6}(m=1)\ F_1\ F_2  \ ,\  c_6(m=1)=\ 6\,c_1\,c_2\,...c_{5}\ ,
\end{equation}
where
\[
 c_k \ \equiv \ 2^k\,{(k!)}^2 \ ,\ k=1,2,3,4,5\ ,
\]
and
\[
F_1 \ = \ V_6^2\ ,
\]
which is the square of the volume of a 6-vertex polytope given by Cayley-Menger determinant, and
\[
F_2 \ = \  \bigg[ 25\,V_6^2\, \bigg(3600\,{V_6^2}\ -\ 48\,\tilde V_5^2\,\tilde V_2^2\ -\
  6\,\tilde V_4^2\,\tilde V_3^2 \
 +\ \frac{1}{9}\,{\big(\tilde V_3^2\big)}^2\,\tilde V_2^2 \bigg)
\]
\[
\ +\ {\tilde V_5^2} \bigg(4\,\tilde V_5^2 {\big(\tilde V_2^2\big)}^2\ -\
    \,{\tilde V_4^2}\,{\tilde V_3^2}\,{\tilde V_2^2} \ +
    \ 54\,{\big(\tilde V_4^2\big)}^2 \bigg) \bigg]\ \geq\ 0.
\]

It is worth noting that
\begin{itemize}
  \item ${\big(\tilde V_2^2\big)}$ contains 15 terms
  \item ${\big(\tilde V_3^2\big)}$ contains 20 terms
  \item ${\big(\tilde V_4^2\big)}$ contains 15 terms
  \item ${\big(\tilde V_5^2\big)}$ contains  6 terms
\end{itemize}
Putting $V_6^2=0$ in $F_2$ we recover \emph{the functional form} of the five-body determinant $D_5(m=1)$.

Seemingly, vanishing $V_5^2=0$ implies $V_6^2=0$.

\end{document}